\documentstyle[aps,preprint,eqsecnum]{revtex}
\makeatletter
                \def\@preprint{}
                \def\preprint#1#2#3#4{%
                \ifpreprintsty
                \def\@preprint{
                \noindent \hbox{#1}\hfill\hbox{#2}\\
                \hbox{#3}\hfill\hbox{#4} \vskip 8pt}%
                \fi
                }
\makeatother

\begin{document}

\preprint{hep-ph/9701413}{CERN-TH/97-6}{}{NTZ 5/97}

\title{Non-local generalization of the axial anomaly and
$x$-dependence of the anomalous gluon contribution}
\author{
        D. M\"uller
        }
\address{TH Division, CERN, 1211 Geneva 23, Switzerland
        \footnote{
			On leave from the Institut f\"ur Theoretische Physik,
			Universit\"at Leipzig, 04109 Leipzig, Germany}
        }
\author{
        O. V. Teryaev
        }
        \address{
        Bogoliubov Laboratory of Theoretical Physics, \\
        Joint Institute for Nuclear Research, 141980 Dubna, Russia
        }

%\date{\today}
\maketitle

\begin{abstract}
The generalization of the axial anomaly is considered. It is shown that
bilocal axial quark operators on the light cone possess, beside the
point-like anomaly, also a light-like anomaly. The consequences for the
definition of anomaly-free quark distribution functions and the effect of
both the gluon coefficient function and splitting kernels in polarized deep
inelastic scattering are discussed.
\end{abstract}

\newpage
\narrowtext

\section{Introduction}

In the last years much effort was put, on both experimental and theoretical
sides, into understanding the spin structure of the proton (for reviews of
the EMC spin crisis, see \cite{review}). New data for the polarized deep
inelastic scattering (DIS) structure function $g_1$ from SLAC
\cite{SLACdatapolDIS} and the SMC \cite{SMCdatapolDIS} are consistent with
the previous EMC \cite{EMCdatapolDIS} result. Two different scenarios can
explain the data, large anomalous sea-quark or large anomalous gluon
interpretation \cite{EfrTer88,AltRos88,CarColMue88}, which are connected
with the problem of the anomalous gluon contribution to $g_1$. In fact, a
renormalization group transformation can shift the anomalous contribution to
the quark sector or to the gluon sector. In this way, either the anomaly
manifested in the matrix element of the axial quark current is responsible
for the large polarization of the sea or it can be perturbatively taken into
account in the gluon coefficient function, while the quark matrix element is
anomaly-free. The dependence of this renormalization group transformation on
the gluon momentum fraction $x$ is still controversial.

{}From the physical point of view, it is more natural that the anomaly is
attributed to the gluons. The gluonic coefficient function then absorbs only
the short-distance contributions, while the quark distribution corresponds
to a conserved operator and may be invoked in the low-energy description of
the proton. The corresponding picture of the nucleon spin structure is most
popular and the gluon polarization is considered to be its most important
unknown ingredient. The problem of its $x$-dependence is the ``physical''
counterpart of the above-mentioned renormalization group transformation.

The accounting for the anomaly by the finite renormalization 
transformation is effectively resulting in the 
substitution in the partonic expression for the structure function
$g_1$,
\begin{equation}
\Delta q \to \Delta \tilde q = \Delta q-{\alpha_s\over {2 \pi}}\Delta g,
\end{equation}
for the first moment of the spin-dependent quark distribution for
each flavour. The oversimplified ``naive" $x$-dependent analogue
is just 
\begin{equation}
\Delta \tilde q(x) = \Delta q(x)-{\alpha_s\over {2 \pi}}\Delta G(x).
\end{equation}
However, such an expression is not compatible, in principle,
with the fact that the gluon contribution starts at one-loop level only.
The simplest, consistent approach was explored in Ref.\ \cite{karl} a few
years ago, by taking the infrared (IR) finite part $E_{\rm IR}$ of the box
diagram, responsible for the photon--gluon interaction:
\begin{equation}
\Delta \tilde q(x) =
	\Delta q(x)-
	{\alpha_s\over {2 \pi}}\int_x^1 {dz \over z}
	E_{\rm IR}\left({x \over z}\right) \Delta g(z).
\end{equation}

However, the choice of the finite part is
ambiguous and the specific role of the axial anomaly remains unclear. This
is why the attempt followed in Ref.\ \cite{ot89} to perform the
decomposition of the box diagram, referring to its kinematical structure.
The anomaly then contributes to the singlet structure, associated with the
spin structure function $g_T=g_1+g_2$ of the photon-gluon 
scattering\footnote{The recent analysis
\cite{rav} confirms that the zero first moment of this structure function is 
manifested
only for regularization schemes, providing the zero moment of $g_1$ as
well.}: 
\begin{equation}
g_T^n={1 \over {2n(n+1)}} \Delta g^n.
\end{equation}

Attributing the anomalous gluon contribution to this structure
function resulted in its $1-x$-dependence. Later, this expression was
obtained and exploited in different approaches \cite{bass,Cheng}. However,
both derivations and interpretations of this result do not seem complete and
some more solid ground is desirable.

The anomalous gluon contribution plays an exceptional role in the general
classification of perturbative QCD contributions. Formally, it is a part of
the next-to-leading order (NLO) contribution. However, due to the well-known
growth of the first moment of the spin-dependent gluon distribution,
compensating one power of $\alpha_s$ makes it essential also at leading
order. The recently calculated two-loop, spin-dependent anomalous
dimensions \cite{MerNee95,Vog96} make it now possible to perform the
complete NLO analysis of the experimental data
\cite{GluReyStrVog96,BalForRid95,GehSti96,WeiMel96}. These NLO calculations
were performed in the dimensional regularization using the 't
Hooft--Veltman--Breitenlohner--Maison (HVBM) scheme \cite{HooVelBreMai} in
which $\gamma^5$ does not anticommute in the unphysical space-time
dimensions. In this minimal subtraction (MS) scheme  both chiral invariance
in the non-singlet sector and the one-loop character of the singlet axial
anomaly are explicitly broken and their restoration requires an additional
finite renormalization.

As the HVBM scheme is used, a further renormalization group transformation
is performed \cite{BalForRid95,For95} in order to make the result compatible
with the standard factorization prescription and the low-energy intuitive
description of the proton. However, only the value of the first moment was
fixed in that procedure, so that there remains an ambiguity in this
transformation. Note that the function $1-x$ was also proposed for this
purpose \cite{Cheng}.

In the present article, we are suggesting the non-local generalization of
the axial anomaly based on the canonical Ward identities for light-ray
operators. This allows us to give the natural description of both
anomaly-free singlet quark distribution and anomalous gluon contribution,
which fix the $x$-dependence of the renormalization group transformation.
The final result is a rigorous proof of the $1-x$-behaviour of the anomalous
contribution in NLO. Here, we do not address the issue of higher loop
corrections to the generalized axial anomaly.

The paper is organized as follows. Section II discusses the chiral invariance
breaking of flavour non-singlet light-ray operators due to the
renormalization, and their restoration by a finite renormalization. Then we
derive Ward identities for light-ray operators and compute the non-local
singlet axial anomaly, which can be expressed as a divergence from a
generalized topological current. In Section III we use our results to define
the anomaly-free singlet quark distribution and show the consequences for
coefficient functions and evolution kernels in NLO approximation.

\section{Axial anomaly of light-ray operators}

As mentioned before, chiral invariance is broken in the HVBM scheme;
however, as it is known from the renormalization of the axial current, it
can be restored by a finite renormalization, so that the non-singlet Ward
identity will be fulfilled. This finite renormalization constant $z^{\rm
NS}$ can be computed for massless QCD from the requirement that the
anticommutativity of $\gamma^5$ is effectively restored \cite{GorLar86},
i.e.
\begin{eqnarray}
\label{finitelocal}
z^{\rm NS}\left\langle
		\left[j^{5,a}_{\mu}\right]\psi \bar{\psi}
	\right\rangle
=
	\left\langle
		\left[j^{a}_{\mu}\right]\psi \bar{\psi}
	\right\rangle  \gamma^5,
\quad \mbox{with}\quad z^{\rm NS}=1-{\alpha_s\over\pi} C_F + O(\alpha_s^2),
\end{eqnarray}
where
$j^{5,a}_{\mu}=
		{1\over 2} \bar{\psi}[\gamma_{\mu},\gamma^5]\lambda^a\psi$
with $\gamma^5=i\gamma^0\gamma^1\gamma^2\gamma^3$
and
$j^{a}_{\mu}=\bar{\psi}\gamma_{\mu}\lambda^a\psi $
are the axial vector and the vector current, respectively,
$\lambda^a$ is a flavour matrix,
the symbol $[...]$ denotes minimal subtraction, and $C_F={4\over 3}$ is
the usual QCD colour factor. {}Flavour and colour indices are suppressed for
simplicity.

The one-loop character of the singlet axial anomaly can also be ensured by
some finite renormalization of the current. This required a two-loop
calculation of the Ward identities sandwiched between the two-gluon state.
The $\alpha_s$ correction to the corresponding $z$ factor,
$z^{\rm S}=1-{\alpha_s\over\pi} C_F +
O(\alpha_s^2)$, coincides with the non-singlet factor $z^{\rm NS}$. Because
of the appearance of so-called light-to-light subdiagrams, this coincidence
is spoiled beyond the one-loop level. Note that the one-loop approximation
of  $z^{\rm S}$ can also be calculated by the requirement that the singlet Ward
identities are fulfilled for the two-quark state.  Since at leading order the
axial anomaly does not appear in this Ward identity, it follows that the
finite renormalization factor is the same as that for the non-singlet channel.

The same problems as discussed above also occur for composite operators on
the light-cone, which appear in the definition of spin-dependent quark
distribution functions. {}For technical reasons we start with a more general
definition of bi-local operators, which are not necessarily on the
light-cone:
\begin{eqnarray}
\label{defopxy}
O_{\mu}^{5,a}(x,y) &=& {1\over 2}
	\bar{\psi}(x) U_s(x,y)\left[\gamma_{\mu},
	\gamma^5\right]\lambda^a \psi(y),
\\
U_s(x,y) &=& P \exp\left\{-ig \int_0^1 d\tau A_k^\mu(x\tau + y[1-\tau])
			t_k(x_\mu-y_\mu)\right\}.
\nonumber
\end{eqnarray}
Here, $U_s(x,y)$ ensures gauge invariance, where the gauge field $A^\mu_k$ is
path-ordered along a straight line connecting the fermion fields. For
light-like distances, i.e. $(x-y)^2=0$, these operators can be expanded in
terms of local twist-2 and twist-3 operators. We set $x=\kappa_1 \tilde{x},
y=\kappa_2 \tilde{x}$, where $\tilde{x}$ is a light-cone vector and, after
contraction with $\tilde{x}^\mu$, we get the leading twist-2 light-ray
operators \cite{DitGeyHorMueRob94}\footnote{ Since $\tilde{x}$ is an
external four-vector that can be kept four dimensional in the
dimensional-regularized operator vertex, it follows that in the HVBM scheme
the relation $\left[\not\!\tilde{x},\gamma^5\right]=
2\not\!\tilde{x}\gamma^5$ is valid, so that indeed Eq.\ (\ref{defop}) comes
from the definition (\ref{defopxy}).}:
\begin{eqnarray}
\label{defop}
O^{5,a}(\kappa_1,\kappa_2;\tilde{x})
	&=& \tilde{x}^\mu O_{\mu}^{5,a}(\kappa_1\tilde{x},\kappa_2\tilde{x}),
\nonumber\\
	&=&
	\bar{\psi}(\kappa_1\tilde{x})
			U(\kappa_1\tilde{x},\kappa_2\tilde{x})
	\not\!\tilde{x}\gamma^5\lambda^a \psi(\kappa_2\tilde{x}).
\end{eqnarray}

To determine the finite non-singlet renormalization constant in the forward
case, we require (for massless QCD), in analogy to Eq.\ (\ref{finitelocal}),
the validity of
\begin{eqnarray}
\label{finitenonlocal}
\int_0^1 dx\;
z^{\rm NS}(x) \left\langle
	\left[O^{5,a}(0,\kappa x;\tilde{x})\right]\psi \bar{\psi}
\right\rangle =
\left\langle
	\left[O^{a}(0,\kappa;\tilde{x})\right]\psi \bar{\psi}
\right\rangle  \gamma^5,
\end{eqnarray}
where $O^{a}$ is analogous to the definition of $O^{5,a}$ in Eq.\
(\ref{defop}),
but without $\gamma^5$ matrix. In the dimensional regularization using the
HVBM scheme, the leading-order result (restricted to the forward case) is
\begin{eqnarray}
\label{resfinitenonlocal}
z^{\rm NS}(x) = \delta(1-x) - {\alpha_s\over\pi} 2C_F (1-x) + O(\alpha_s^2),
\end{eqnarray}
while in the Pauli--Villars regularization, chiral invariance holds true
without finite renormalization.

As in the case of the axial current one expects, in leading order, that the
same finite renormalization as in Eq.\ (\ref{resfinitenonlocal}) has to be
performed for the singlet light-ray operator. Beyond the leading order this
is no longer true and the question is: How can we fix this finite
renormalization constant?

After we have seen that the restoration of chiral invariance requires
an  $x$-dependent finite renormalization, a second question arises: Does the
axial anomaly also depend on $x$? To answer this question we use the
equation of motion to derive Ward identities for the divergence of
the non-local operators (\ref{defopxy}).
A straightforward calculation provides that the divergence of
this operator is
\begin{eqnarray}
\label{WI1}
\left(\partial_x^\mu + \partial_y^\mu\right)
O_{\mu}^{5,a}(x,y) &=&
	\Omega^{\rm EOM}(x,y) +
	 i\bar{\psi}(x)U_s(x,y)\gamma^5 m^a \psi(y)-
\nonumber\\
	& & ig \int_0^1 d\tau\;
	\bar{\psi}(x)U_s(x,x\tau)
	\gamma_{\alpha} F^{\alpha\beta}(x\tau+y[1-\tau]) (x_\beta-y_\beta)
	 \times
	\nonumber\\
	& &\qquad U_s(y[1-\tau],y)\gamma^5\lambda^a \psi(y),
	\\
\label{WI2}
\Omega^{\rm EOM}(x,y) &=&
	\bar{\psi}(x)\big(
					[\not\!\!D(x)-i m] U_s(x,y)+U_s(x,y)[\not\!\!D(y)-i m]
				\big)
	\gamma^5\lambda^a \psi(y),
\end{eqnarray}
where $\Omega^{\rm EOM}(x,y)$ denotes the equation of motion operators,
$F_{\alpha\beta} = F^a_{\alpha\beta}  t^a$ is the field strength tensor,
and $m_{ij}^a=(m_i+m_j)\lambda^a_{ij}$ is a mass matrix.

{}For $(x-y)^2 \not= 0$ the operator is ultraviolet-finite and the Ward
identity (\ref{WI1}) should be satisfied. However, the situation will be
changed if we go to a light-like (not only short, as usually discussed)
distance. Then the operator has to be renormalized, and anomalous terms can
occur. For instance, if we naively anticommute $\not\!\!D(y)$ with $\gamma^5$
in the second expression in the r.h.s.\ of Eq.\ (\ref{WI2}) then the use of
the equation of motion will provide only contact terms. However, in the HVBM
scheme the non-anticommutativity of $\gamma^5$ gives an anomalous term,
which should be cancelled by a finite renormalization (see the discussion in
\cite{Colren}), so that the anticommutativity of $\gamma^5$ is effectively
restored as discussed above.
We calculated this anomalous term and after an appropriate definition of the
appearing operators the finite renormalization constant
(\ref{resfinitenonlocal}) was extracted from the Ward identity. As discussed
above for the local case, because of the absence of the singlet axial anomaly
in the Ward identities for the Green functions of
quark fields at one-loop order, it follows that $z^{\rm S}(x)$ coincides with
$z^{\rm NS}(x)$ at this order.

{}For the singlet case ($a=0, \lambda^0_{ij}=\delta_{ij},
m_{ij}^0=2m_i\delta_{ij}$), we expect an anomalous term of the form
$\epsilon^{\alpha\beta\gamma\delta}
		F_{\alpha\beta}(\cdots) F_{\gamma\delta}(\cdots) $,
which can be computed from the difference of the l.h.s.\ and r.h.s.\ of the
Ward identity (\ref{WI1}) sandwiched between the two gluon states (see
diagrams in Fig.\ 1). We now set $x=\kappa_1 \tilde{x}, y=\kappa_2
\tilde{x}$ with $\tilde{x}^2=0$ and choose the light-cone gauge, i.e.
$\tilde{x}A=0$. After a straightforward calculation, it turns out that in
both Pauli--Villars regularization and HVBM scheme the same anomaly appears,
so that the singlet Ward identity is actually given by
\begin{eqnarray}
\label{WIan1}
\left(\partial_{(\kappa_1 \tilde{x})}^\mu +
\partial_{(\kappa_2 \tilde{x})}^\mu\right)
O_{\mu}^{5,0}(\kappa_1 \tilde{x},\kappa_2 \tilde{x}) &=&
%i\bar{\psi}(\kappa_1 \tilde{x})\gamma^5 m^0 \psi(\kappa_2 \tilde{x})
O_F(\kappa_1,\kappa_2;\tilde{x}) +
{\alpha_s\over 4\pi} N_f \int_0^1 dx \int_0^{1-x} dy\; 2\times
\nonumber\\
& &\hspace{-1cm}
F^a_{\mu\nu}\left([\kappa_1(1-x)+\kappa_2 x]\tilde{x}\right)
 \tilde{F}^{a\mu\nu}\left([\kappa_2(1-y)+\kappa_1 y]\tilde{x}\right),
\\
& &\hspace{-6.6cm}\mbox{where}
\nonumber\\
\label{WIan1defOF}
O_F(\kappa_1,\kappa_2;\tilde{x}) &=&
- ig \int_{\kappa_1}^{\kappa_2} d\tau\;
	\bar{\psi}(\kappa_1 \tilde{x})
		\gamma_{\alpha} F^{\alpha\beta}(\tau\tilde{x}) \tilde{x}_\beta
 	\gamma^5  \psi(\kappa_2 \tilde{x}).
\end{eqnarray}
Here, $\tilde{F}^{a\mu\nu} =
	{1\over 2}\epsilon^{\mu\nu\alpha\beta} F^a_{\alpha\beta}$,
with $\epsilon_{0123}=+1$, $\alpha_s = {g^2 \over 4\pi}$, and $N_f$ is the
number of flavours. Here we applied the equation of motion and neglected the
quark masses. In the local case, which follows from setting
$\kappa_1=\kappa_2=\kappa$, the operator
$O_{\mu}^{5,0}(\kappa\tilde{x},\kappa\tilde{x})$ coincides with the local
singlet axial current $j_{\mu}^{5,0}(\kappa\tilde{x})$ and from Eq.\
(\ref{WIan1}) we recover the well-known expression for the local anomaly:
${\alpha_s\over 4\pi} N_f F^a_{\mu\nu}\left(\kappa\tilde{x}\right)
\tilde{F}^{a\mu\nu}\left(\kappa\tilde{x}\right)$.

Finally, we introduce a non-local generalization of the topological current
\begin{eqnarray}
\label{topocurrent}
K^{\mu}(x,y)&=& {\alpha_s\over 4 \pi} N_f\epsilon^{\mu\alpha\beta\gamma}
	\left[
		A^a_{\alpha}(x) \partial_\beta A^a_{\gamma}(y)
	- {g\over 3} f_{abc} A^a_{\alpha}(x) A^b_{\beta}(y) A^c_{\gamma}(y)+
	\left\{x \leftrightarrow y\right\}\right],
%\nonumber\\
%	   & &\hspace{2,1cm}A_a^{\alpha}(y) \partial_x^\beta A_a^{\gamma}(x)
%	- {g\over 3} f_{abc} A_a^{\alpha}(y) A_b^{\beta}(x) A_c^{\gamma}(x)
%	\big],
\end{eqnarray}
with $x=\kappa_1 \tilde{x}$ and $y=\kappa_2 \tilde{x}$, so that the anomaly
in Eq.\ (\ref{WIan1}) can be written as a divergence
\begin{eqnarray}
\label{divtopocurrent}
\left(
	\partial^\mu_{(\kappa_1 \tilde{x})}+\partial^\mu_{(\kappa_2 \tilde{x})}
\right)
	K_{\mu}(\kappa_1 \tilde{x},\kappa_2 \tilde{x})
&=&
	{\alpha_s\over 4 \pi} N_f
	F^a_{\mu\nu}\left(\kappa_1 \tilde{x}\right)
	\tilde{F}^{a\mu\nu}\left(\kappa_2 \tilde{x}\right).
\end{eqnarray}
We are now able to define the anomaly-free operator
\begin{eqnarray}
\label{anofreeop}
\tilde{O}_{\mu}^{5,0}(\kappa_1 \tilde{x},\kappa_2 \tilde{x}) &=&
	O_{\mu}^{5,0}(\kappa_1 \tilde{x},\kappa_2 \tilde{x}) -
\nonumber\\
& &	2\int_0^1 dx \int_0^{1-x} dy\;
			K_{\mu}([\kappa_1(1-x)+\kappa_2 x]\tilde{x},
					[\kappa_2(1-y)+\kappa_1 y]\tilde{x}).
\end{eqnarray}

\section{Consequences for polarized distribution functions}

As shown in the previous Section light-ray operators possess anomalous
contributions. For the non-singlet case these anomalies are a pure artefact
of choosing a non-invariant chiral renormalization scheme. The polarized
quark distribution functions should be defined in a chiral invariant manner,
so that in a general renormalization scheme an additional finite
renormalization is necessary
\begin{eqnarray}
\label{defquarkNS}
\Delta q^{\rm NS}(x,Q^2)= \int_x^1 {dy\over y}\; z^{\rm NS}_5(y)
\int {d\kappa \over 2\pi (\tilde{x}S)}
					\left\langle P,S\left|
\left[O^{5,{\rm NS}}(0,\kappa;\tilde{x})\right]
					 \right|P,S \right\rangle
				e^{i (x/y)\kappa (\tilde{x}P)},
\end{eqnarray}
where $S^\rho$ denotes the polarization vector of the nucleon and the
renormalization point square $\mu^2$ is set equal to the momentum transfer
squared $Q^2$. The leading-order approximation of $z^{\rm NS}_5(y)$ in the
HVBM scheme is given in Eq.\ (\ref{resfinitenonlocal}). This finite
renormalization affects the NLO approximation of both quark coefficient
functions and splitting kernels and agrees with the additional
renormalization group transformation performed in the NLO calculation
\cite{MerNee95,Vog96}.

Because of the anomaly, the naive definition of the singlet distribution
function
\begin{eqnarray}
\label{defnaiveSquark}
\Delta \tilde{\Sigma}(x,Q^2)&=&
	\sum_{i=u,d,...} \Delta q_i(x,Q^2)+\Delta\bar{q}_i(x,Q^2),\qquad
\Delta\bar{q}_i(x,Q^2)=\Delta q_i(-x,Q^2)
\nonumber\\
	&=&\int_x^1 {dy\over y}\; z^{\rm S}_5(y)
		\int {d\kappa \over 2\pi (\tilde{x}S)}
					\left\langle P,S\left|
\left[O^{5,0}(0,\kappa;\tilde{x})\right]+\left\{\kappa\to -\kappa \right\}
					 \right|P,S \right\rangle
				e^{i (x/y)\kappa (\tilde{x}P)}
\end{eqnarray}
(please note that $\Delta\tilde{\Sigma}$ is defined in terms of the operator
$O^{5,0}$) cannot be interpreted as probability for finding a polarized
quark flavour singlet configuration with given longitudinal momentum
fraction $x$. In addition to the finite multiplicative
renormalization\footnote{This renormalization is due to the diagrams without
two-gluon intermediate states, which are the same for the non-singlet and
the singlet case, so that $z^{\rm S}_5(y)=z^{\rm NS}_5(y)$, as discussed
in Section II.}, it
is also necessary to remove the axial anomaly from the definition of this
function \cite{EfrTer88,AltRos88,CarColMue88}. Thus, one has to define the
singlet distribution function in terms of the anomaly-free operator
(\ref{anofreeop}), which provides
\begin{eqnarray}
\label{defSquark}
\Delta\Sigma(x,Q^2) &=& \Delta\tilde{\Sigma}(x,Q^2) - k(x,Q^2),
\\
& &\hspace{-2.4cm}\mbox{where}
\nonumber\\
\label{defanomcont}
k(x,Q^2) &=&
2 \int_x^1 {dy\over y} (1-y)\int {d\kappa \over 2\pi (\tilde{x}S)}
			\left\langle P,S\left|
				\tilde{x}^{\mu} K_{\mu}(0,\kappa\tilde{x})
				+\left\{\kappa\to -\kappa \right\}
			\right|P,S \right\rangle
					e^{i (x/y)\kappa (\tilde{x}P)},
\end{eqnarray}
and the generalized topological current $K_{\mu}$ is defined in
Eq.\ (\ref{topocurrent}).

The problem that $K_{\mu}$ is gauge-variant, and thus that $k(x,Q^2)$
contains also unphysical components, can somehow be resolved by the choice
of a physical gauge. In the light-cone gauge, the gauge-invariant twist-2
gluon operator
\begin{eqnarray}
\label{defgluonOp}
G(\kappa_1,\kappa_2;\tilde{x})=
 i	\tilde{x}^\alpha \tilde{F}_{a\alpha\beta}(\kappa_1\tilde{x})
	F_a^{\beta\gamma}(\kappa_2\tilde{x}) \tilde{x}_\gamma
\end{eqnarray}
can be expressed in the forward case by
\begin{eqnarray}
\label{conglonOpandK}
G(0,\kappa;\tilde{x})=
	-\left({\alpha_s\over 2\pi} N_f\right)^{-1} i\tilde{x}\partial_{\kappa}
	\tilde{x}^\mu K_{\mu}(0,\kappa\tilde{x}).
\end{eqnarray}
Furthermore, from the definition of the gluon distribution function
\begin{eqnarray}
\label{defgluon}
\Delta g(x,Q^2)={1\over x(\tilde{x}P)} \int {d\kappa \over 2\pi (\tilde{x}S)}
					\left\langle P,S\left|
							G(0,\kappa;\tilde{x})
					+\left\{\kappa\to -\kappa \right\}
					 \right|P,S \right\rangle
							e^{ix\kappa (\tilde{x}P)},
\end{eqnarray}
and from Eqs.\ (\ref{defanomcont}) and (\ref{conglonOpandK}), and after
performing a partial integration, we find that $k(x,Q^2)$ is actually given
in terms of the gauge-invariant gluon distribution function. Thus, to remove
the axial anomaly of the polarized quark distribution function it is
sufficient to subtract a certain amount of the gluon distribution function:
\begin{eqnarray}
\label{defSquarkfin}
\Delta\Sigma(x,Q^2) = \Delta \tilde{\Sigma}(x,Q^2) -
						K(x)\otimes \Delta g(x,Q^2),
%\nonumber\\
\quad
K(x)= -{\alpha_s\over \pi} N_f (1-x),
\end{eqnarray}
where the convolution is defined as 
\begin{eqnarray}
\label{convolution}
A(x)\otimes B(x) = \int_0^1dy\int_0^1 dz\; \delta(x-y z) A(y) B(z).
\end{eqnarray}

Indeed, removing the axial anomaly is equivalent to the following
(additive) renormalization group transformation:
\begin{eqnarray}
\label{rengroptrans}
\Delta C_g(x) &=& \Delta \tilde{C}_g(x) + K \otimes\Delta\tilde{C}_q(x) ,
\quad\Delta C_q(x) = \Delta \tilde{C}_q(x),
\\
\Delta P_{gg}(x) &=& \Delta\tilde{P}_{gg}(x) +
					K \otimes\Delta\tilde{P}_{gq}(x),
\quad
\Delta P_{gq}(x) = \Delta \tilde{P}_{gq}(x),
\nonumber\\
\Delta P_{qq}(x) &=& \Delta\tilde{P}_{qq}(x) -
					K \otimes\Delta\tilde{P}_{gq}(x) ,
\nonumber\\
\Delta P_{qg}(x) &=& \Delta\tilde{P}_{qg}(x) - {\beta\over g} K(x)
		+ K\otimes [\Delta\tilde{P}_{qq}-\Delta\tilde{P}_{gg} -
					\Delta\tilde{P}_{gq}\otimes K](x),
\nonumber
\end{eqnarray}
where $\Delta C_i$ are the coefficient functions, $\Delta P_{ij}$ are the
spin-dependent splitting kernels,
and $\beta=\mu{d\over d\mu} g(\mu)$ is the renormalization group coefficient
of the running coupling constant.

Because of gauge invariance the operator product analysis suggests that the
zero moment of $\Delta C_g$ vanishes. After the transformation
(\ref{rengroptrans}) is performed, the gauge-variant axial anomaly
contributes to the gluonic sector, so that the zero moment is now given by
\begin{eqnarray}
\label{0momentg}
\int_0^1 dx\; \Delta C_g(x) =
\int_0^1 dx\;  K(x) + O\left(\alpha_s^2\right)=
- {\alpha_s\over 2 \pi} N_f + O\left(\alpha_s^2\right).
\end{eqnarray}
{}For completeness we give the difference of the splitting kernels in NLO:
\begin{eqnarray}
\label{changekernel}
\Delta P_{gg}(x)-\Delta\tilde{P}_{gg}(x)  &=&
					-\left(\Delta P_{qq}(x)-\Delta \tilde{P}_{qq}(x)\right)
\nonumber\\
	&=& \left({\alpha_s\over 2\pi}\right)^2 2 C_F N_F
		\left[3(1-x)+(2+x)\ln x\right],
\nonumber\\
\Delta P_{qg}(x)- \Delta \tilde{P}_{qg}(x)  &=&
			\left({\alpha_s\over 2\pi}\right)^2 N_f
		\big\{
				C_F\left[(1-x)(1-4\ln(1-x) + 2\ln x)\right]+
\nonumber\\
&&\hspace{20,5mm}
				C_A\left[(1-x)(-16+4\ln(1-x)) -4(2+x)\ln x\right]
		\big\},
\end{eqnarray}
where the Casimir operator $C_A$ is equal to the number of colours.
Note, that these differences will vanishes in the limit $x\to\ 1$.
For small $x$ they contribute only to the subleading behaviour of the NLO
result (the leading terms are given by $\ln^2x$).

Another opportunity to perform the evolution coincides with the one proposed
by Cheng \cite{Cheng}. Namely, one should perform the evolution in the
gauge-invariant (say, $\overline{\mbox{MS}}$) scheme and {\it afterwards}
restore the anomaly-free distribution by applying Eq.\ (\ref{defSquarkfin}).
The $1-x$ behaviour in this approach is actually coming from the mass term 
in the box graph. The contact with our derivation may be achieved 
by noting that the cancellation of normal and anomalous divergence,
resulting in the effective conservation of axial current in the 
limit of infinite quark mass, is valid for the non-local anomaly
as well. It is especially clear for the Pauli-Villars regularization,
when the contribution of regulator fermions (calculated in Section II)
is looking, up to the sign, precisely such us that of the quark masses, which
is the starting point of the approach of Cheng. 

{}For practical purposes, irrespective of the used evolution scheme,
it is possible to define an effective gluon distribution, which is just the
combination appearing in $g_1$, i.e. the convolution of $(1-x)$ with
$\Delta g(x)$:
\begin{equation}
\label{defeff}
\Delta g^{\rm eff}(x, Q^2)= 2(1-x)\otimes \Delta g(x,Q^2),
\end{equation}
so that the first moments of effective gluon distribution coincide with the
``original'' one, while the structure function $g_1$ has at leading order
the simple partonic form, suitable for the extraction of partonic
distributions from the experimental data \cite{soffer}:
\begin{eqnarray}
\label{defg1}
g_1(x,Q^2) =
{1 \over 2}\Delta \tilde{\Sigma}(x,Q^2)=
{1 \over 2}\left[ \Delta \Sigma(x,Q^2) -
{\alpha_s\over {2 \pi}} N_f \Delta g^{\rm eff}(x,Q^2)\right].
\end{eqnarray}

It is convenient to have the evolution equation directly for effective
distribution functions. While the diagonal kernels are not changed, the
moments of the off-diagonal kernels are changed in a straightforward manner:
\begin{eqnarray}
\label{effkernel}
\Delta P_{gq}^{\rm eff}(n)={2\over {n(n+1)}}\Delta{P}_{gq}(n),
\qquad
\Delta P_{qg}^{\rm eff}(n)={n(n+1)\over 2}\Delta {P}_{qg}(n).
\end{eqnarray}
Note, especially, that the influence of the effective gluon distribution to
the quark evolution, governed by $\Delta P_{qg}^{\rm eff}$, appears to be
much less singular in $n$. We are not presenting the explicit form for
effective NLO anomalous dimensions, which are rather lengthy. At leading
order, when the $x$ dependence is easily restored,
Eq.\ (\ref{effkernel}) results in the equations
\begin{eqnarray}
\label{efflead}
\Delta P_{gq}^{\rm eff}(x)={\alpha_s\over \pi}C_F[3(x-1)-(x+2)\ln x],
\quad
\Delta P_{qg}^{\rm eff}= -
{\alpha_s\over 4\pi}N_f {d \over{dx}}\delta(1-x).
\end{eqnarray}

The first moments of the splitting kernels provide an important check for the
normalization of the non-local contribution and determine the evolution
in the corresponding sum-rules. So we summarize the consequences for the zero
moment of the splitting kernels, which come from general renormalization
arguments of the axial vector current and the topological current, which are
verified up to two- and three-loop order, respectively \cite{Lar93}.
  From current conservation of
$j^{5,0}_\mu=\tilde{j}^{5,0}_\mu-k_\mu$ (here, $\tilde{j}_\mu^{5,0}$ refers
to the original definition in terms of quark fields),
the Adler--Bardeen theorem, and the renormalization properties of
gauge-invariant operators, it follows that
\begin{eqnarray}
\label{0momentqq}
\int_0^1 dx\; \Delta P_{gg}(x) &=&
\int_0^1 dx\; \Delta \tilde{P}_{gg}(x)+\gamma_j = -{\beta\over g} +\gamma_j,
\nonumber\\
\int_0^1 dx\; \Delta P_{gq}(x) &=&
\int_0^1 dx\; \Delta\tilde{P}_{gq}(x)=-{\gamma_j\over N_f[\alpha_s/(2\pi)]},
\nonumber\\
\int_0^1 dx\; \Delta P_{qg}(x) &=&\int_0^1 dx\; \Delta\tilde{P}_{qg}(x)=0,
\nonumber\\
\int_0^1 dx\; \Delta P_{qq}(x) &=&
\int_0^1 dx\; \Delta\tilde{P}_{qq}(x) -\gamma_j =0,
\end{eqnarray}
where the three-loop order approximation for the anomalous dimension of the
axial vector current $\gamma_j$ and for the $\beta$-function are
known\footnote{
Recently, the four-loop approximations for the $\beta$-function
and for $\bar{\gamma}_j$ were calculated. Here, $\bar{\gamma}_j$ denotes the
anomalous dimension in the MS scheme and the finite renormalization constant
$z^{\rm S}$ is known up to the order $\alpha_s^2$. To obtain $\gamma_j$ and,
therefore, also the first moment of $\Delta P_{gq}(x)$ at three-loop
order it is necessary to know $z^{\rm S}$ up to the order $\alpha_s^3$. As
proposed above, this $\alpha_s^3$ correction can be obtained by the
requirement that the singlet Ward identities for pure quark Green functions
are fulfilled at three-loop order. The four-loop calculation of both sides
of the Ward identities for the two-gluon state gives then the consistency
check for the Adler-Bardeen theorem.}
\cite{Lar93,beta3}:
\begin{eqnarray}
\label{axcuranomdim}
\gamma_j &=&
-\left({\alpha_s\over 4\pi}\right)^2 6 C_F N_f +
 \left({\alpha_s\over 4\pi}\right)^3
	\left(-{142\over 3}C_A + {4\over 3} N_F + 18 C_F
	\right)C_F N_f,
\\
\label{beta}
{\beta\over g}&=&
	-{\alpha_s\over 4\pi}
		\left({11\over 3} C_A -{2\over 3}N_f\right)
	-\left({\alpha_s\over 4\pi}\right)^2
			\left({34\over 3} C_A^2-2 C_f N_f -{10\over 3} C_A N_f\right)-
\nonumber\\
	&& \left({\alpha_s\over 4\pi}\right)^3
		\left(
			{2857\over 54} C_A^3 + C^2_F N_f-
			{205\over 18} C_F C_A N_f -{1415\over 54} C_A^2 N_f +
			{11\over 9} C_F N_f^2 + {79\over 54} C_A N_f^2
		\right).
\end{eqnarray}

These results allow \cite{EfrTer90} the extraction of the anomaly equation
renormalization for a number of loops exceeding that of $\gamma_j$ by 1. For
the leading three-loop contribution, this coincides with the calculation of
Anselm and Johansen \cite{AnsJoh}, while the four-loop correction require a
one-loop finite correction to the gluon matrix element of the topological
current \cite{Zakharov}. As a result, the four-loop correction takes the
form
\begin{eqnarray}
\label{renanom}
Z_j =
-\left({\alpha_s\over 4\pi}\right)^3 6 C_F N_f^2 +
 \left({\alpha_s\over 4\pi}\right)^4
	\left(-{214\over 3}C_A + {4\over 3} N_F + 18 C_F
	\right)C_F N_f.
\end{eqnarray}

\acknowledgements

We wish to thank V. Braun, B. Geyer, S. Larin, E. Leader, D. Robaschik
and G. Rudolph for valuable discussions. O. V. T. would like to thank the
Naturwissenschaftlich-Theoretisches Zentrum and the Graduiertenkolleg
``Quantenfeldtheorie" at Leipzig University for the kind hospitality during
his stay, where substantial parts of this work were done. D.M. was
financially supported by the Deutsche Forschungsgemeinschaft. O.V.T. was
supported by a research grant of the Saxonian Ministry of Science and
Culture.

%\newpage

%\section*{Figure caption}

%\noindent
%Diagrammatical representation for the l.h.s.\ (a) and r.h.s.\ (b) of the
%non-local singlet Ward identity (\ref{WIan1}) in the one-loop approximation.
%The symbol $dO$ refers to the divergence
%$\left(\partial_{(\kappa_1 \tilde{x})}^\mu +
% \partial_{(\kappa_2 \tilde{x})}^\mu\right)
%O_{\mu}^{5,0}(\kappa_1 \tilde{x},\kappa_2 \tilde{x}) $
%and $O_F$ denotes the operator $O_F(\kappa_1,\kappa_2;\tilde{x})$ defined
%in Eq.\ (\ref{WIan1defOF}).

\begin{thebibliography}{99}
\bibitem{review}
E. Reya,
in QCD - 20 Years Later, eds. P. Zerwas and H.A. Kastrup
(World Scientific, Singapore, 1993), p. 272;
S.J. Brodsky,
in Proc. of the XXI Summer Institute on Particle Physics,
SLAC-Report-444, p. 81;
%eds. L. DePorcel and C. Dunwoodie (SLAC,1994)
M. Anselmino, A. Efremov and E. Leader, Phys. Rep. 261 (1995) 1.
\bibitem{SLACdatapolDIS}
E143 Collaboration, K. Abe et al., Phys. Rev. Lett. 74 (1995) 346,
	ibid. 75 (1995) 25.
\bibitem{SMCdatapolDIS}
SM Collaboration, D. Adams et al., Phys. Lett. B329 (1994) 399,
	ibid. B357 (1995) 248.
\bibitem{EMCdatapolDIS}
EM Collaboration, D. Adams et al., Phys. Lett. B206 (1988) 364.
\bibitem{EfrTer88}
A.V. Efremov and O.V. Teryaev,
%JINR Dubna report E2-88-287,
in Proc. of the Symposium on Hadron Interactions - Theory and Phenomenology,
(Bechyne, Czechoslovakia, 1988), p. 302.
\bibitem{AltRos88}
G. Altarelli and G.G. Ross, Phys. Lett. B212  (1988) 391.
\bibitem{CarColMue88}
R.D. Carlitz, J.C. Collins and A.H. Mueller, Phys. Lett. B214 (1988) 229.
\bibitem{karl}
 J. Ellis, M. Karliner and C.T. Sachrajda, Phys. Lett. B231 (1989) 497.
\bibitem{ot89}
A.V. Efremov and O.V. Teryaev, SPIN-89, in Proc. of III Serpukhov
International Workshop, p. 77;
J. Soffer and O.V. Teryaev, Phys. Rev. D51 (1995) 25.
\bibitem{rav}
P. Mathews, V. Ravindran and K. Sridhar, TIFR-TH-96-40,
hep-ph/9607385.
\bibitem{bass}
S.D. Bass,
Phys. Lett. B312 (1993) 345.
\bibitem{Cheng}
H.Y. Cheng, Int. J. Mod. Phys. A11 (1996) 5109.
\bibitem{MerNee95}
R. Mertig and W.L. van Neerven, Z. Physik C70 (1996) 637.
\bibitem{Vog96}
W. Vogelsang,
Phys. Rev. D59 (1996) 2023,
Nucl. Phys. B475 (1996) 47.
\bibitem{GluReyStrVog96}
M. Gluck, E. Reya, M. Stratmann and W. Vogelsang,
Phys. Rev. D53 (1996) 4775.
\bibitem{BalForRid95}
R.D. Ball, S. Forte and G. Ridolfi,
Phys. Lett. B378 (1996) 255.
\bibitem{GehSti96}
T. Gehrmann and W.J. Stirling,
Phys. Rev. D53 (1996) 6100.
\bibitem{WeiMel96}
T. Weigl and W. Melnitchouk, Nucl. Phys. B465 (1996) 267.
\bibitem{HooVelBreMai}
G. 't Hooft and M. Veltman, Nucl. Phys. B44 (1972) 189;
P. Breitenlohner and \\ D. Maison, Commun. Math. Phys. 52 (1977) 11.
\bibitem{For95}
S. Forte,
CERN preprint CERN-TH/95-305, hep-ph/9511345.
\bibitem{GorLar86}
S.G. Gorishny and S.A. Larin, Phys. Lett. B172 (1986) 109.
\bibitem{DitGeyHorMueRob94}
F-M. Dittes, B. Geyer, J. Ho\v{r}ej\v{s}i, D. M\"uller and D. Robaschik,
Fortschr. Phys. 42 (1994) 101.
\bibitem{Colren}
J.C. Collins, Renormalization (Cambridge U.P., Cambridge, 1984), section 13.
\bibitem{soffer}
C. Bourelly, F. Buccella, O. Pisanti, P. Santorelli and J. Soffer,
CPT-96-PE-3327, hep-ph/9604204.
\bibitem{Lar93}
S.A. Larin,
Phys. Lett. B303 (1993) 113.
\bibitem{beta3}
O.V. Tarasov, A.A. Vladimirov and A.Yu. Zharkov, Phys. Lett. 93B
(1980) 429; \\
S.A. Larin and J.A.M. Vermaseren,
Phys. Lett. B303 (1993) 334.
\bibitem{EfrTer90}
A.V. Efremov and O.V. Teryaev, Sov. J. Nucl. Phys. 51 (1990) 943.
\bibitem{AnsJoh}
A.A. Anselm and A.A. Johansen, JETP Lett. 49 (1989) 214.
\bibitem{Zakharov}
M.A. Shifman and A.I. Vainshtein, Nucl. Phys. B277 (1986) 456;
A.I. Vainshtein and V.I. Zakharov, Sov. Phys. JETP 68 (1989) 701.
\end{thebibliography}
\end{document}